\documentclass[a4paper,10pt]{article}
\usepackage{graphicx}

\begin{document}

%\preprint{}

\title{Cluster size entropy in the Axelrod model of social influence:
small-world networks and mass media}\author{Y\'{e}rali Gandica $^{1}$\\
%EndAName
\emph{Instituto Venezolano de Investigaciones Cient\'{\i}ficas.}\\
\emph{Centro de F\'{\i}sica, Altos de Pipe, Carretera Panamericana, Km 11,}\\
\emph{Caracas 1020A, Venezuela.} \and A. Charmell, J. Villegas-Febres \\
%EndAName
\emph{Departamento de Estado S\'{o}lido, Instituto de F\'{\i}sica,}\\
\emph{Grupo de Quimicof\'{i}sica de Fluidos y Fen\'{o}menos} \\
\emph{Interfaciales (QUIFFIS), Departamento de Qu\'{i}mica,}\\
\emph{Universidad de Los Andes, M\'{e}rida 5101, Venezuela} \and I. Bonalde \\
%EndAName
\emph{Departamento de F\'{\i}sica, Universidad Sim\'{o}n Bol\'{\i}var,}\\
\emph{Valle de Sartenejas, Baruta, }\\
\emph{Apartado Postal 89000, Caracas 1080-A, Venezuela.}}
\maketitle

\begin{abstract}
We study the Axelrod's cultural adaptation model using the
concept of \textit{cluster size entropy} $S_{c}$ that gives
information on the variability of the cultural cluster size
present in the system. Using networks of different topologies,
from regular to random, we find that the critical point of the
well-known nonequilibrium monocultural-multicultural
(order-disorder) transition of the Axelrod model is
unambiguously given by the maximum of the $S_{c}(q)$
distributions. The width of the cluster entropy distributions
can be used to qualitatively determine whether the transition
is first- or second-order. By scaling the cluster entropy
distributions we were able to obtain a relationship between the
critical cultural trait $q_c$ and the number $F$ of cultural
features in regular networks. We also analyze the effect of the
mass media (external field) on social systems within the
Axelrod model in a square network. We find a new partially
ordered phase whose largest cultural cluster is not aligned
with the external field, in contrast with a recent suggestion
that this type of phase cannot be formed in regular networks.
We draw a new $q-B$ phase diagram for the Axelrod model in
regular networks.
\end{abstract}
\footnotetext{$^{1}$Correspondence author. E-mail: ygandica@gmail.com}
%\pacs{89.75.Fb, 87.23.Ge, 89.75.Hc}
%\keywords{Axelrod, mass media, external field, randomness, entropy}

%\maketitle
\newpage \baselineskip1.0cm
\section{Introduction}

Recently, research in complex systems has paid particular
attention to elucidate some of the mechanisms leading to
interesting economic and social phenomena, such as opinion
formation, self-organization, distribution of richness,
formation of coalitions, land and air traffics, and evolution
of social structures \cite{anderson,oliveira,castellano2}.
Several theoretical approaches have been proposed to understand
social systems. Axelrod \cite{axelrod} introduced a model to
study in particular the dissemination of cultures among
interacting individuals or agents in which a) the more
culturally similar are the agents, the greater the chance of
interaction between them, and b) interaction increments
similarity between individuals. Among the interesting results
obtained with this model is a nonequilibrium transition from a
monocultural state, where all agents share the same cultural
features, to a multicultural state, where individuals mostly
have their own features, as the cultural diversity increases
\cite{castellano1}.

The Axelrod model has been widely employed to analyze the
effect of cultural drift caused by noise
\cite{klemm1,klemm2,sanctis1}, repulsive interactions between
individuals \cite{radillo1}, and mass media
\cite{gonavella2,gonavella3,gonavella1,rodriguez1,peres1} on
social systems. The effect of the mass media, which is normally
modeled as a uniform external field with values in the range
[0,1], has been of special interest since it was discovered
that in finite square networks there is a critical field
$B_c=0.05$ above which the state of the system is always
multicultural (disordered) \cite{gonavella2}. It was also found
that for values of the number of cultural traits $q>q_c$ the
state is always multicultural, independently of the field value
$B$ \cite{gonavella2}. Here, $q_c$ is the critical value of the
nonequilibrium multicultural-monocultural phase transition in
the absence of an external field ($B=0$). Later, it was shown
that even for vanishing $B$ the monocultural state is
destabilized in very large systems \cite{peres1}. More
recently, it was found that for the Axelrod model in complex
networks the system can order in a vector state different from
the one imposed by the external field \cite{gonavella1}. All
these are very counterintuitive findings, opposed to what is
classically found in physics for spin systems that monotonously
align with the external field. Interestingly, ordered phases
that are not aligned with the external field are only
encountered in fully connected, in random and in scale-free
networks. It was thus claimed that long-range interactions,
absent in regular lattices, are required for the appearance of
this result \cite{gonavella1}.

The nonequilibrium phase transition (hereafter we will omit the
term "nonequilibrium") of the Axelrod model is characterized by
an order parameter $\phi$ that is usually defined as the
average size of the largest cultural cluster $C_{max}$ in the
system normalized by the total number of agents $N$;
$\phi=C_{max}/N$. In the monocultural (ordered) state
$\phi\rightarrow1$ and in the multicultural (disordered) state
$\phi\rightarrow0$. Even though this order parameter
appropriately identified both the ordered and the disordered
phase emerging in the Axelrod model, it does not clearly define
the critical region. Here, we show that the cluster size
entropy $S_c$, which is defined in terms of the probability
that an occupied site of the lattice belongs to a cluster
containing $s$ sites, can be used as another powerful tool for
the analysis of the phenomenology of the Axelrod model. The
cluster size entropy measures the number of clusters of
different sizes and is related to the diversity of the system
\cite{tsang2}. Theoretically, as a function of the probability
of occupation $S_c$ should be zero in both the ordered and the
disordered phase, since the former is constituted by a single
cluster of the size of the system and the latter is formed by a
large number of small clusters of similar sizes. $S_c$ should
have a maximum at the transition where the diversity of cluster
sizes is maximum. Thus, a peak develops as the phase transition
takes place, which leads to a much better definition of the
critical region.

In statistical physics, cluster size entropy has been used in
the study of problems such as percolation
\cite{SA,HK,tsang2,tsang3} and complex systems
\cite{radillo1,gomes,tsang1,tsang2,tsang3}. In their studies on
percolation, Tsang and co-workers \cite{tsang2,tsang3} found
that the cluster entropy shows a maximum at the percolation
threshold, where a group of neighboring occupied sites forms a
cluster that expands from one edge of the 2D lattice to the
opposite one causing an abrupt decrease in the cluster entropy
of the system. In the context of the Axelrod model, cluster
entropy measures the number of cultural groups of different
sizes and was first used by Villegas-Febres and Olivares-Rivas
\cite{jcv} in an attempt to establish a connection with
thermodynamics. Cluster entropy was also utilized within the
Axelrod model to partially characterize the inclusion of
repulsion among agents in a regular lattice \cite{radillo1}.

To demonstrate the usefulness of the cluster entropy in the
analysis of complex networks, here we employ this property for
the first time to carry out a throughout study on the
monocultural-multicultural phase transition of the Axelrod
model. We analyze the effects of the topology of the network
and the mass media on this phase transition. Regarding the
topology, we vary the probability $p$ of random rewiring
between sites from 0 (regular networks) to 1 (random networks).
In addition to determining more exactly the critical value
$q_c$ of the Axelrod model, we establish a mathematical
expression that relates $q_c$ with the important parameter $F$
(number of cultural features) in finite regular lattices and
display in a much clearer form some other known properties.
Considering the imposition of the mass media, we find that
partially ordered states that are not aligned with the field
can be formed in short-range-interaction regular networks, in
contraposition to the claim by Gonz\'{a}lez-Avella \textit{et
al.} \cite{gonavella1}. A new $q-B$ phase diagram is proposed
for the Axelrod model in regular lattices.

\section{Axelrod model}

The original Axelrod model is defined on a square lattice of
$N$ sites (social agents). The state of the $i^{th}$ agent is
defined by a set of $F$ cultural features (e.g., religion,
sport, politics, etc.) represented by a vector
$C_{i}=(C_{i1},C_{i2},...,C_{iF})$. Each feature $C_{ik}$ of
the agent $i$ is first randomly assigned with a uniform
distribution of the integers in the interval $[0,q-1]$. The
variable $q$ defines the cultural traits allowed per feature
and thus measures the cultural variability in the system. There
are $q^F$ possible cultural states.

The procedure to establish the dynamics of the system is as
follows: (1) Choose randomly two nearest neighbor agents $i$
and $j$, then (2) calculate the number of shared features
(cultural overlap) between the agents $\ell_{ij}=\sum_k^F
\delta_{C_{ik},C_{jk}}$. If $0<\ell_{ij}<F$ then (3) pick up
randomly a feature $k$ such that $C_{ik}\neq C_{jk}$ and with
probability $\ell_{ij}/F$ set $C_{ik} = C_{jk}$. These time
steps are iterated and the dynamics stops when a frozen state
is reached; i.e., either $\ell_{ij}=0$ or $\ell_{ij}=F, \forall
i,j$. A cluster is a set of connected agents with the same
state. Monocultural or ordered phases are composed of a cluster
of the size of the system where $\ell_{ij}=F, \forall i,j$.
Multicultural or disordered phases consist of two or more
clusters.

To study the effect of an external field in the original
Axelrod model just described some modifications are needed. We
define a uniform external field as a vector
$M=(m_1,m_2,...,m_F)$, where $m_n \in [0,1,...,q-1]$, with
strength $B\in[0,1]$. This parameter $B$ regulates the
probability for the agent-field interactions. Each agent has a
probability $B$ of interacting with the field and a probability
($1-B$) of interacting with one of its nearest neighbors.

In the dynamic described above the agent $j$ is substituted by
the field $M$ and the whole sequence follows in the same way:
(1) Choose randomly an agent $i$, then (2) with probability $B$
agent $i$ and the field $M$ interact, (3) calculate the number
of shared features (cultural overlap) between the agent and the
field $\ell_{iM}=\sum_k^F \delta_{C_{ik},M_n}$. If
$0<\ell_{iM}<F$ then (4) pick up randomly a feature $k$ such
that $C_{ik} \neq m_k$ and with probability $\ell_{iM}/F$ set
$C_{ik} = m_k$. If agent $i$ and the field $M$ do not interact,
(5) choose randomly an agent $j$ in the nearest neighborhood of
agent $i$. If with probability $1-B$ agents $i$ and $j$
interact, (6) compute the cultural overlap $\ell_{ij}=\sum_k^F
\delta_{C_{ik},C_{jk}}$. If $0<\ell_{ij}<F$ (7) with
probability $\ell_{ij}/F$ set $C_{ik} = C_{jk}$. The dynamics
stops when a frozen state is reached.

To generate randomized lattices, required for the topology
analysis, we used the Watts-Strogatz algorithm \cite{WS}.
Starting with a regular network of $N$ agents (with periodic
boundary conditions), each link is visited and with probability
$p$ is removed and rewired at random (avoiding self-linked
nodes). This random rewiring process produces networks with
topologies that go from perfect regularity $(p=0)$ to full
randomness $(p=1)$.

\section{cluster entropy}

The cluster size entropy is defined as \cite{tsang2}
\begin{equation}
S_{c}(P)=-\sum_s W_s(P)\ln W_s(P) \,, \label{entropy}
\end{equation}
where $P$ is the probability of occupation (probability $1/q$
of taking a particular value of the cultural trait) and
$W_{s}(P)$ the probability that an agent belongs to a cluster
of size $s$.

\section{network topology in the Axelrod model}

%
%------------------------FIG-------------
\begin{figure*}[t]
\scalebox{0.65}{\includegraphics{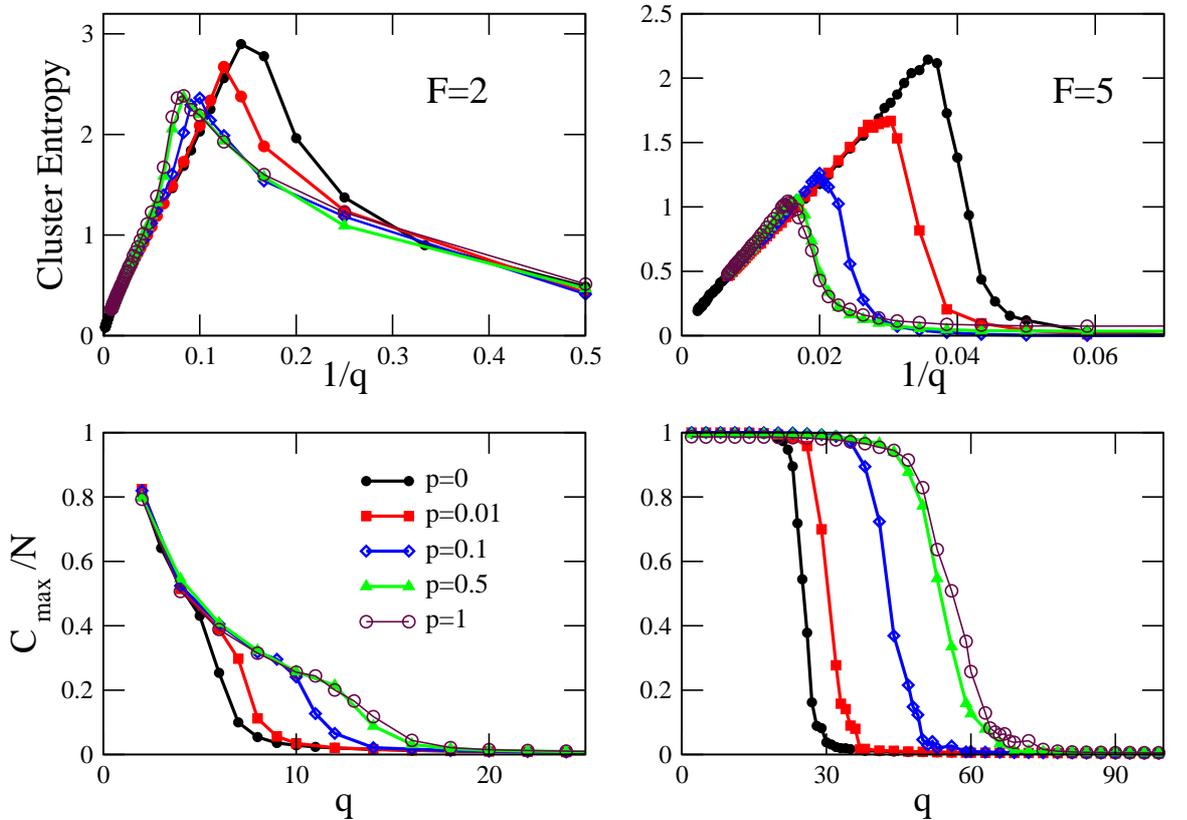}}
\caption{\label{fig:1}{(Color online) Cluster size entropy as a
function of the probability of occupation $1/q$ and order
parameter versus $q$ for $F=2$ and $F=5$ in networks of $40
\times 40$ with disorder parameter $p$ equals $0, 0.01, 0.1,
0.5$ and $1$. The peak of the cluster entropy distributions
occurs at the onset of the phase transitions as characterized
by the order parameter $\phi$ (see text).}}
\end{figure*}
%-------------------------FIG-----------------
%

Here we vary the disorder parameter $p$ of the network from 0
to 1 to see the effect on the monocultural-multicultural phase
transition of the Axelrod model. Figure \ref{fig:1} displays
the cluster size entropy as a function of the probability of
occupation $P=1/q$ and, for comparison, the order parameter
$\phi$ against the cultural trait $q$ for F=2 and F=5 in
networks of $N=40 \times 40$ agents. There are various relevant
issues in this figure that need consideration. The maximum of
the cluster entropy curves is located around the corresponding
onset trait $q_c$ of the monocultural-multicultural phase
transitions as studied with $\phi$. The peak of a cluster size
entropy distribution corresponds to the state of maximum size
disorder that occurs at the transition. Thus, the average size
of the largest cluster becomes finite when the maximum size
disorder occurs. The size entropy brings out the fact that as
the system moves from a disordered phase, in which most agents
occupy each a single cluster, to an ordered phase, in which all
agents occupy just one cluster of the system size, the system
goes through a critical region where agents become agglomerated
in clusters of different sizes. We see in the curves of
Fig.~\ref{fig:1} that the maximum of the cluster entropy
defines more precisely the critical value $q_c$ than the finite
value of the order parameter $\phi$.

Another feature to observe in Fig.~\ref{fig:1} is that the
critical value $1/q_c$ ($q_c$) becomes smaller (larger) as $p$
goes from $0$ (regular networks) to $1$ (fully random
networks), in complete agreement with the results obtained with
the order parameter $\phi$ by Klemm \textit{et al.}
\cite{klemm3}. We believe that the collapse of the curves
corresponding to $p>0.5$ is due to finite-size effects and not
to the lack of any further dynamics that indeed seems to
develop up to $p=1$ in networks of large size \cite{klemm3}.

Next, we point out that the height of the cluster entropy
curves in Fig.~\ref{fig:1} reveals a new characteristics of the
Axelrod model that was impossible to uncover by means of the
order parameter $\phi$. As $p$ increases the maximum of the
size entropies becomes smaller, suggesting less cluster
diversity at the transition. Thus, it appears that in random
networks, due to the presence of long-range links, the
formation of the system-size cluster defining the ordered state
of the Axelrod model is more efficient (explores a smaller
region of the cluster size space) than in regular networks.

\begin{figure}
\scalebox{0.3}{\includegraphics{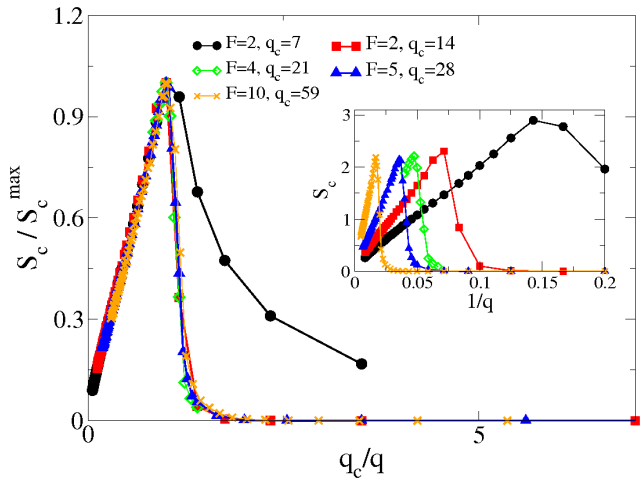}}
\caption{\label{fig:2}{(Color online) Normalized cluster size
entropy $S_c/S_c^{max}$ against the normalized probability of
occupation $q_c/q$ for a network of disorder $p=0$ and size $40
\times 40$ for different values of $F$. The data suggest that
the Axelrod model for $F=2$ is in a different universality
class from that for $F>2$ (see text). The inset shows the
regular data.}}
\end{figure}

\begin{figure}
\vspace{8pt} \scalebox{0.3}{\includegraphics{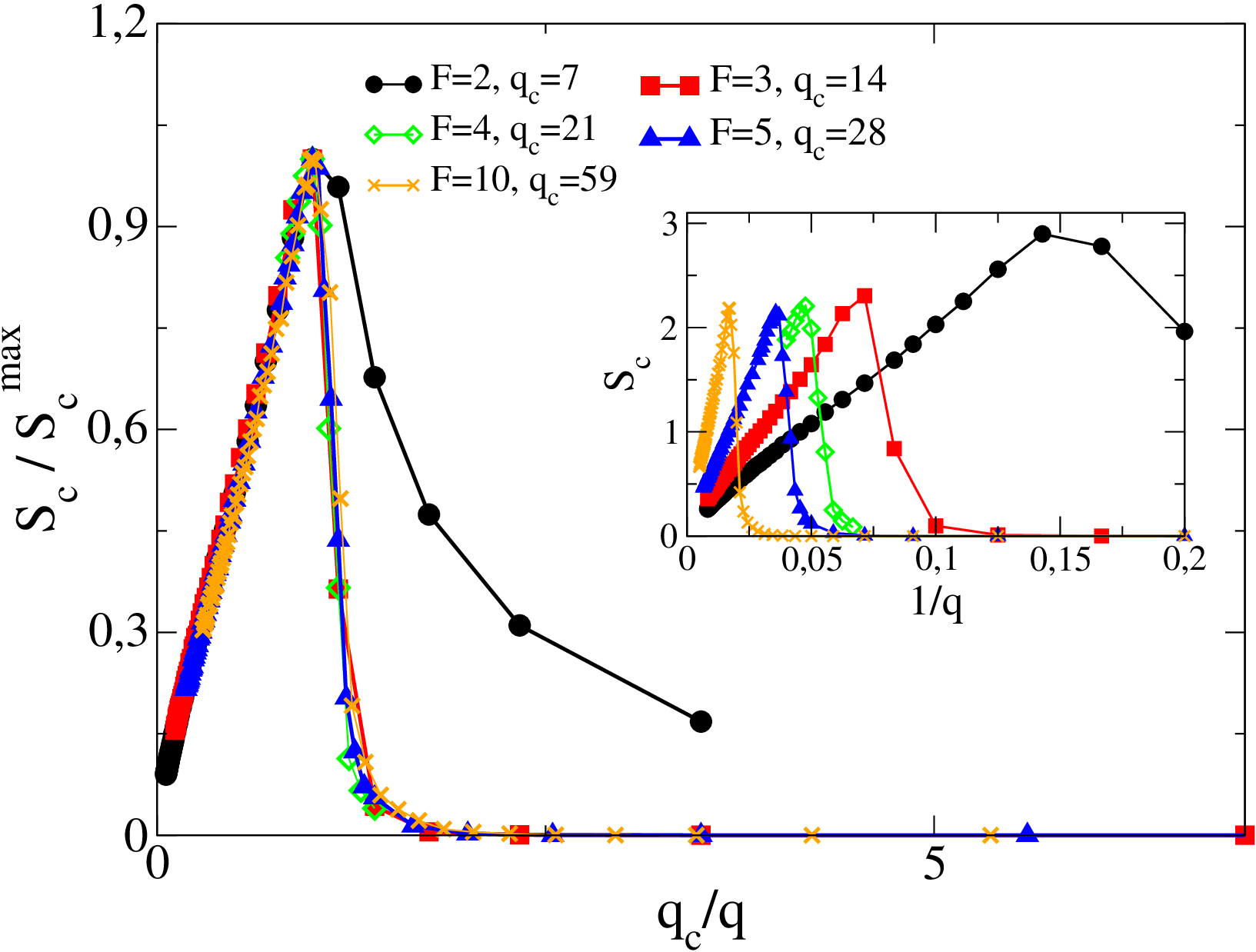}}
\caption{\label{fig:3}{(Color online) Normalized cluster size
entropy $S_c/S_c^{max}$ against the normalized probability of
occupation $q_c/q$ for a network of disorder $p=0.1$ and size
$40 \times 40$ for different values of $F$. The inset shows the
regular data.}}
\end{figure}

\begin{figure}
\scalebox{0.3}{\includegraphics{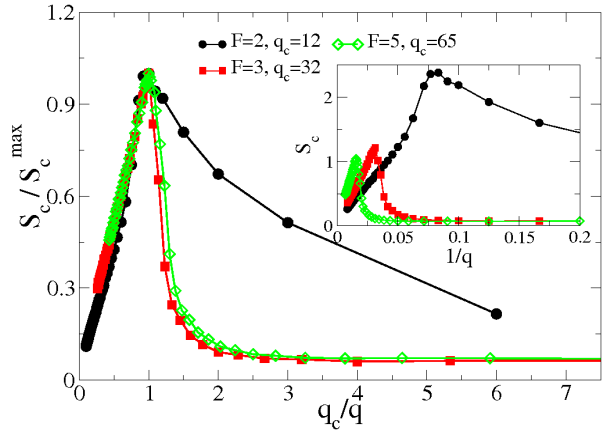}}
\caption{\label{fig:4}{(Color online) Normalized cluster size
entropy $S_c/S_c^{max}$ against the normalized probability of
occupation $q_c/q$ for a network of disorder $p=1$ and size $40
\times 40$ for different values of $F$. The inset shows the
regular data.}}
\end{figure}

The most highlighted findings of the present analysis on the
effect of topology in the Axelrod model are observed in the
scaled data of Figs.~\ref{fig:2}, \ref{fig:3}, and \ref{fig:4}
for network randomness $p=0, 0.1$ and $1$, respectively.
%
%------------------------FIG-------------
%-------------------------FIG-----------------
%
%------------------------FIG-------------
%-------------------------FIG-----------------
%
%
%------------------------FIG-------------
%-------------------------FIG-----------------
These figures display the normalized cluster entropy
$S_c/S_c^{max}$ versus the normalized probability of occupation
$q_c/q$ for different values of the cultural-feature number
$F$. Here, $S_c^{max}$ is the value of the cluster entropy at
the peak maximum. The insets show the corresponding regular
data. All three plots show that, independently of the network
randomness, the overall dynamics of the Axelrod model for $F=2$
is unambiguously different from that for any value $F>2$, and
that the dynamics is the same for all values $F>2$. The plots
indicate that the Axelrod model for $F=2$ and $F>2$ must belong
to different "universality classes" in the terminology of
statistical physics. This agrees with previous works that
claimed that the transition is first-order type for $F>2$ and
second-order type for $F=2$ \cite{castellano1,klemm1}.

%
%------------------------FIG-------------
\begin{figure}[b]
\vspace{8pt} \scalebox{0.3}{\includegraphics{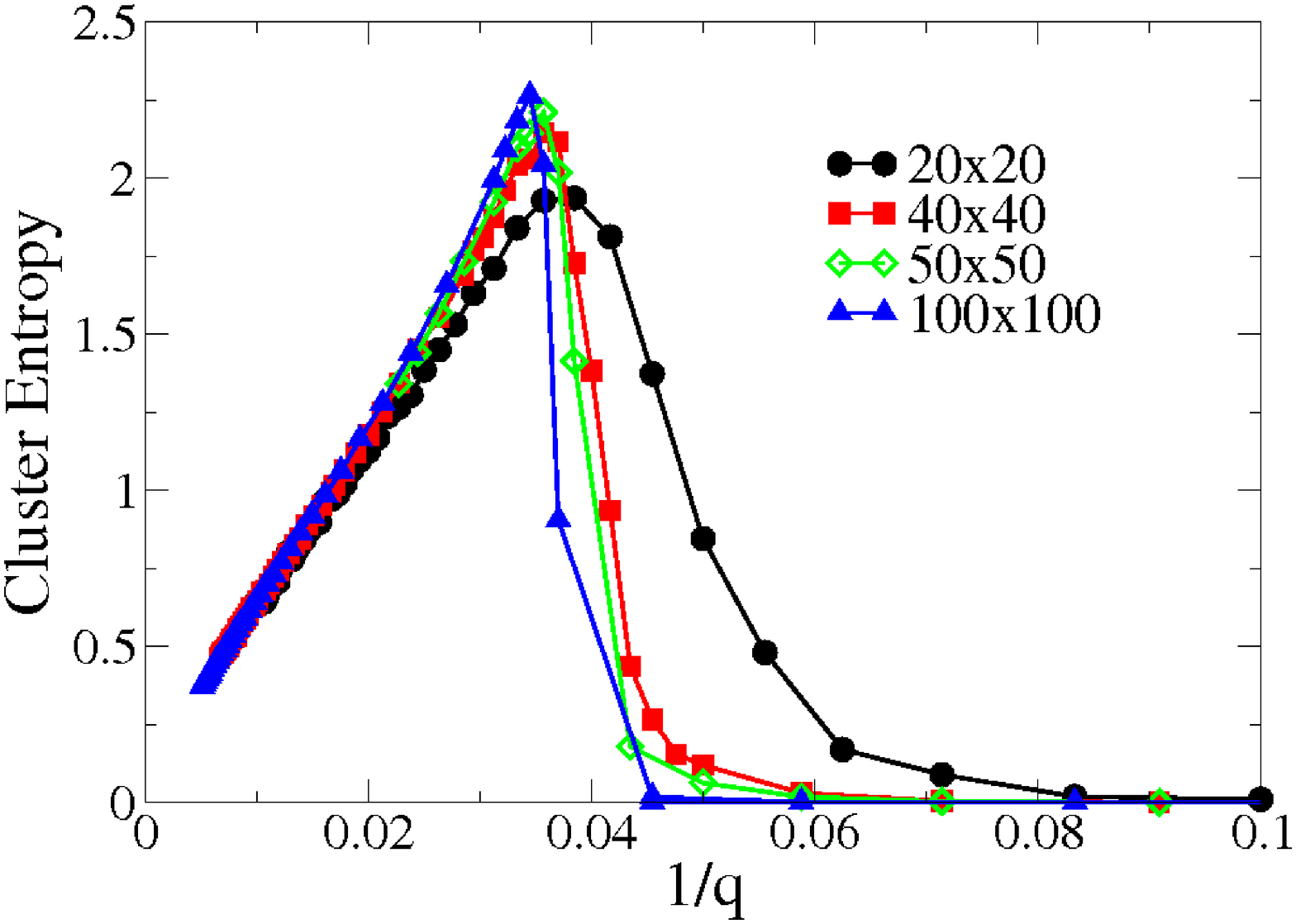}}
\caption{\label{fig:5}{(Color online) Cluster size entropy
versus probability of occupation in regular networks ($p=0$) of
different sizes.}}
\end{figure}
%-------------------------FIG-----------------
%

The width of the cluster size entropy distribution may give a
clue of the order of the transition. The more narrow
distributions of the cluster entropy for $F>2$
(Figs.~\ref{fig:2}, \ref{fig:3}, and \ref{fig:4}) compared with
the distributions for $F=2$ indicate that the system crosses
over from disorder (a large number of clusters exist) to order
(just one cluster size is present) in a sudden manner, which is
typical of first-order-like transitions. Broad distributions,
instead, suggest that the system moves from one regime to the
other in a smooth manner. A finite number of clusters of all
sizes starts to form in the disordered phase and as the system
crosses over to the ordered side those clusters of all sizes
agglomerate in just one of the size of the system. This is the
behavior expected in second-order-like transitions.

%
%------------------------FIG-------------
\begin{figure*}[t]
\scalebox{0.65}[0.62]{\includegraphics{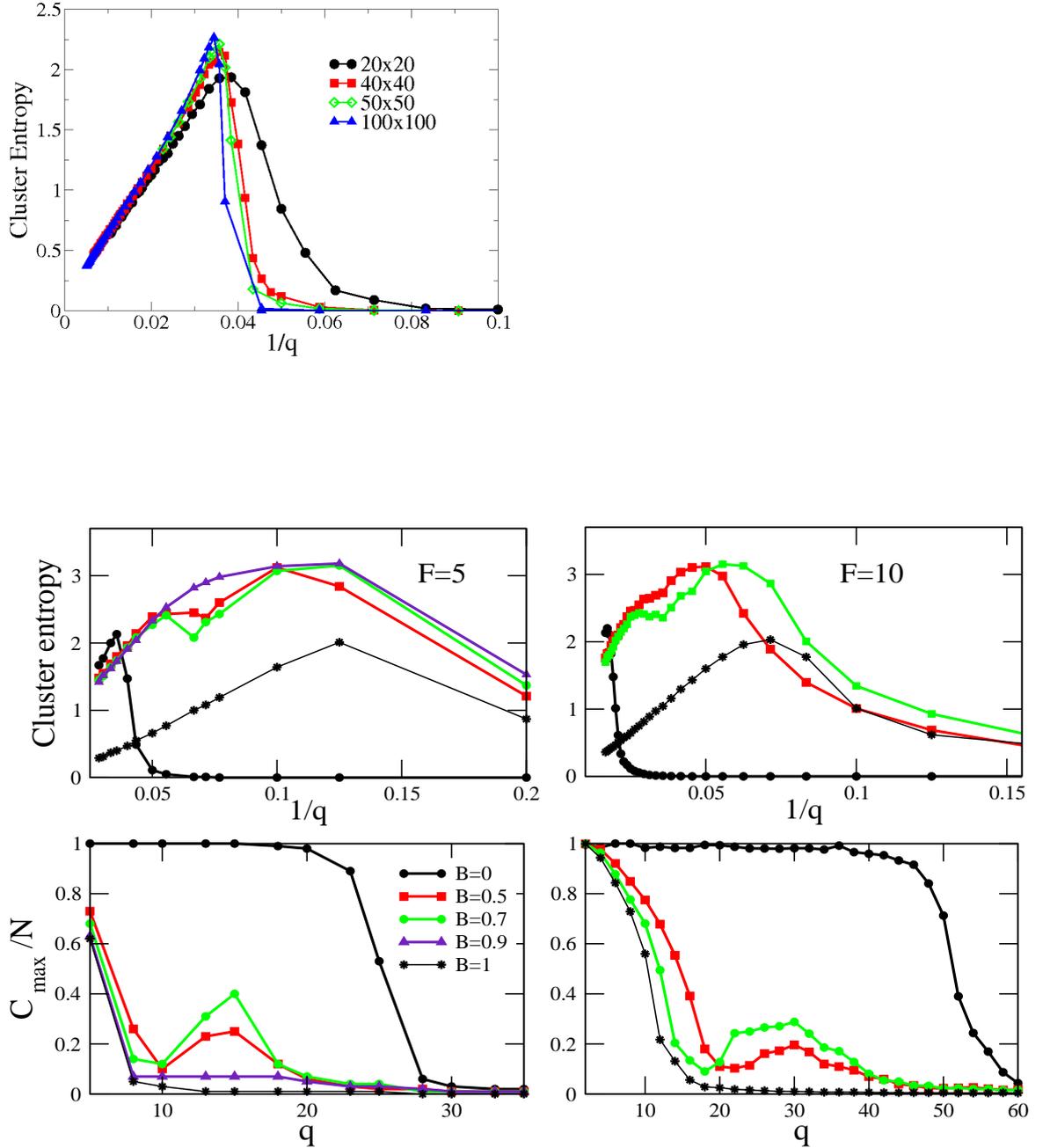}}
\caption{(Color online) Cluster entropy $S_c$ against $1/q$ and
order parameter $\phi$ as a function of $q$ for $F=5$ and
$F=10$ in a $40 \times 40$ regular network for various field
strengths $B$. For both values of $F$ a second transition
develops in the field range $0.2 < B < 0.8$. For the sake of
clarity the case $B=0.9$ not displaying a second transition is
omitted in $F=10$. } \label{fig:6}
\end{figure*}
%-------------------------FIG-----------------
%

Notably, the normalized cluster entropies are independent of
$F$ for $(q_c/q) < 1$, indicating that there is a unique
Axelrod dynamics for values $q>q_c$. This conclusion is robust
against finite-size effects, as is demonstrated by the data
shown in Fig.~\ref{fig:5}.

It is widely known that the monocultural-multicultural phase
transition occurs at a critical value $q_c$ that increases as F
augments \cite{castellano1,klemm4}. However, there is no
expression relating $q_c$ and $F$. Here, we provide such an
equation valid in regular lattices ($p=0$) of finite sizes:
\begin{equation}
q_{c} \approx 7*(F-1) \,. \label{critq}
\end{equation}
This expression 1) defines the value of the cultural trait $q$
above which the system will always be in a multicultural state
and 2) states that for any value of $F$ the system will always
be monocultural for $q < 7$. Equation~(\ref{critq}) appears to
hold for other values of the network randomness $p$ with a
pre-factor that becomes larger as $p$ increases.

\section{external field or mass media in the Axelrod model}

Gonz\'{a}lez-Avella  \textit{et al.} \cite{gonavella1} found
that in the presence of an external field the Axelrod model
displays a rich $q-B$ phase diagram in fully connected, in
random and in scale-free 2D networks for $F=10$. A second
ordered phase was found that is not aligned with the external
field and that in complex networks does not cover the whole
system. Gonz\'{a}lez-Avella  \textit{et al.} argue that this
ordered phase is caused by the long-range interactions
characteristic of complex networks, and that such a phase does
not exist in regular (short-range interaction) networks.

Figure \ref{fig:6} shows the cluster entropy against $1/q$ and
the order parameter versus $q$ in a $40 \times 40$ regular
network for several values of the field strength $B$ and for
$F=5$ and $F=10$. One of the effects of the external field is
to move the critical value $1/q_c$ toward higher occupation
probabilities, as occurs when the number of cultural features
$F$ is reduced (see Figs. \ref{fig:1} and \ref{fig:2}).
Interestingly, the limiting $(1/q_c)=0.128$ as $B\rightarrow 1$
is very close to the limiting $(1/q_c)=0.142$ as $F\rightarrow
2$.

More relevant, a second peak is observed in the cluster
entropies of Fig.~\ref{fig:6} for $0.2 < B < 0.8$ that
indicates the occurrence of a second phase transition. These
cluster entropies are constituted by two overlapping entropy
distributions whose peaks correspond to the onset of phase
transitions. The second peaks in $S_c$ in the upper panels of
Fig.~\ref{fig:6} correspond to reentrant behaviors in the order
parameter $\phi$ in the lower panels. That is, as the value of
$q$ is lowered the system moves from a multicultural to an
ordered-like phase, then returns to the multicultural phase,
and finally goes to a monocultural phase. The ordered-like
phase makes the phase diagram of the Axelrod model in regular
networks more complex than previously thought
\cite{gonavella1}. The appearance of the second transition both
for $F=5$ and for $F=10$ indicates that its existence does not
depend on $F$.

A further analysis reveals that the system in the extra ordered
phase, called here "crossing phase", is mainly formed by a
large cultural cluster whose state is not aligned with the
external field. This crossing phase, whose order parameter
$\phi<1$ (meaning that the largest cluster does not have the
size of the system), is closely related to the "orthogonal"
ordered phase reported by Gonz\'{a}lez-Avella \textit{et al.}
\cite{gonavella1} of the Axelrod model in $50 \times 50$ random
networks. Contrasting with the conclusions by
Gonz\'{a}lez-Avella \textit{et al.}, our results indicate that
long-range interactions are \textit{not} fully required in the
Axelrod model to form an ordered state whose orientation is not
parallel to the external field.

%
%------------------------FIG-------------
\begin{figure}
\scalebox{0.48}{\includegraphics{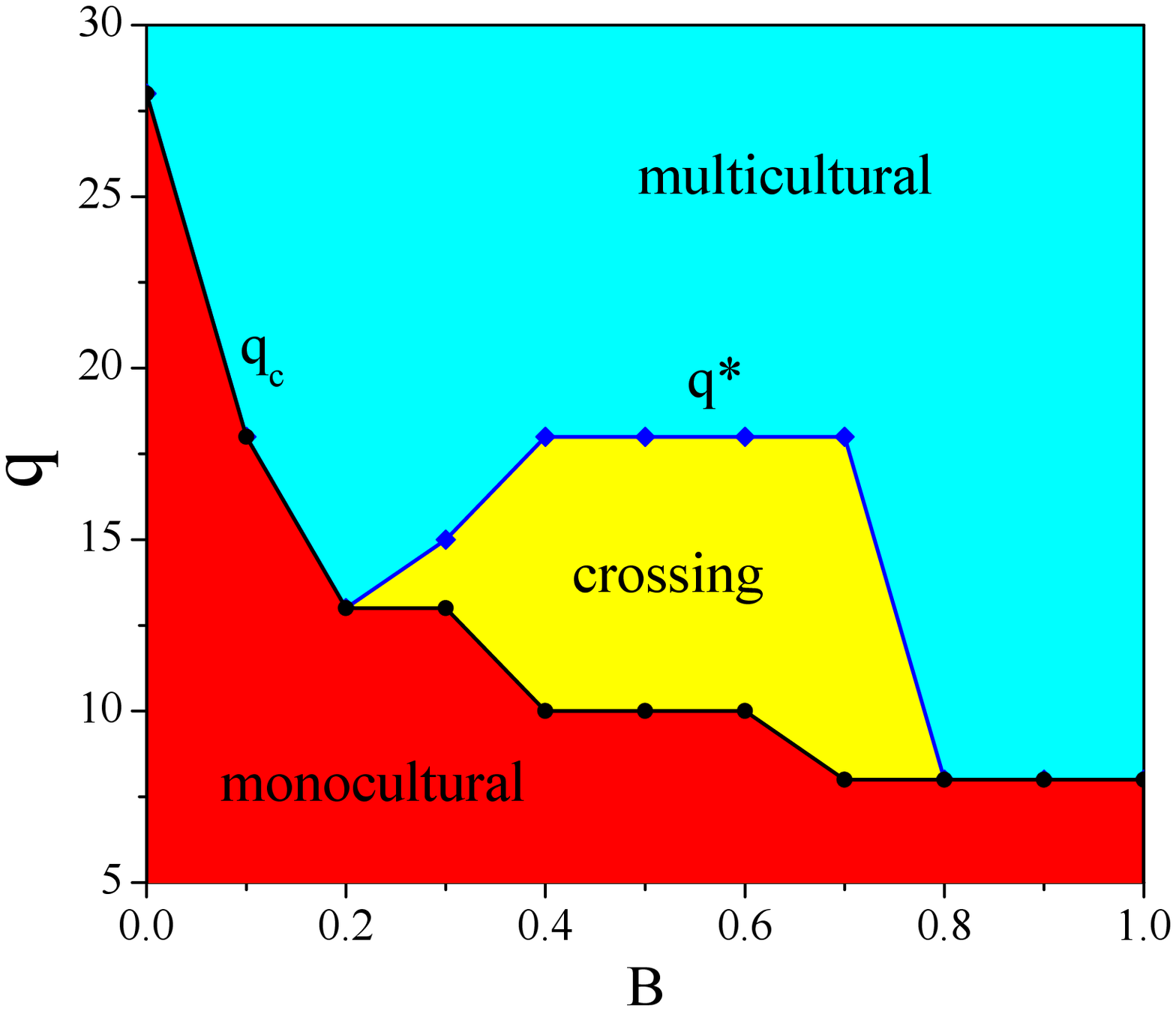}}
%\scalebox{0.7}{\includegraphics{fig7.eps}}
\caption{\label{fig:7}{(Color online) $q-B$ phase diagram of
the Axelrod model obtained for $F=5$ in a $40 \times 40$
regular network. Each point represents an average of 50-100
realizations.}}
\end{figure}
%-------------------------FIG-----------------
%
In Fig.~\ref{fig:7} we draw a new phase diagram of the Axelrod
model in regular networks by using the peaks of the size
entropy distributions for $F=5$. The highest peak at $q_c$ of
each $S_c$ distribution in Fig.~\ref{fig:6} agrees with the
onset of the monocultural-multicultural transition in the
corresponding $\phi$ curve. The second peak at $q^*$ in the
entropy distributions is assigned to the multicultural-crossing
transition. In the monocultural phase (colored red) the state
($\phi \sim 1$) is aligned with the external field for $q <
q_c$. In the crossing phase (colored yellow) the ordered state
($\phi < 1$) is not parallel to the external field for $q_c < q
< q^*$ and $0.2 < B < 0.8$. In this region there is no overlap
between the ordered state and the field. The multicultural
phase (colored blue) is completely disordered. In particular,
the state in regular networks for $F=5$ is always monocultural
for $q < 8$ and multicultural for $q > 28$ independently of the
external field, in agreement with Eq.~(\ref{critq}).

Finally, in connection with the analysis of the previous
section, we remark here that the data for $F=10$ in
Fig.~\ref{fig:6} provide another good example to show that the
cluster size entropy can indicate better the critical value
$q_c$ than the order parameter. One can see that for the cases
$B=0.5$ and $B=0.7$ the transition onset is better determined
by the second peak of the cluster size entropy than by the
initial finite value of the order parameter, although this last
property displays quite well the transition.

\section{Summary}

Thus far, the cluster size entropy has been hardly used in the
analysis of phase transitions. Here, we showed that the cluster
size entropy $S_c$ is a valuable tool that can be utilized as a
complement of the order parameter $\phi$. Using the cluster
entropy, we were able to both reproduce most of the results
previously known for the Axelrod model in square networks and
find new relevant results. We showed by a simple analysis that
the Axelrod model for $F=2$ and $F>2$ belong to a different
"universality" class. For regular lattices it was determined an
expression that relates $q_c$ and $F$ and that defines the
asymptotic values of the trait $q$ for the presence of
multicultural and monocultural phases in the system.

We found a new partially ordered phase for the Axelrod model in
regular lattices, in which the vector state of the largest
cultural cluster is not aligned with the external field or mass
media. This phase is similar to one previously reported for
fully connected, scale-free and random networks, and leads to a
new cultural trait-field ($q-B$) phase diagram for the Axelrod
model in regular networks. The results suggest that long-range
interactions are not completely necessary for the existence of
an ordered state that is not oriented along the applied field.

\section{acknowledgments}

J.V-F thanks CDCHTA for Grants C-1449-07-05-F and ADG-C-09-95.
Y.G. acknowledges support from Misi\'{o}n Ciencia and IVIC
doctoral fellowships. We thank computational support from
Leonardo Trujillo at IVIC and Centro Nacional de C\'{a}lculo
Cient\'{\i}fico (CeCalCULA) of the Universidad de Los Andes.

%\bibliography{sociophysics}

\end{document}